\documentclass[preprint,floats,aps,prl,eqnum,showpacs,column]{revtex4}
\usepackage{color,graphicx}
\usepackage{amsfonts}
\usepackage{amssymb}

\begin{document}

\title{Can we distinguish between black holes and wormholes by their Einstein-ring systems?}

\author{Naoki Tsukamoto,\footnote{Electronic address:11ra001t@rikkyo.ac.jp}
Tomohiro Harada and Kohji Yajima
}

\affiliation{
Department of Physics, Rikkyo University, Tokyo 171-8501, Japan 
}
\date{\today}

\begin{abstract}

For the last decade, gravitational lensing in the strong gravitational field has been studied eagerly. 
It is well known that, for the lensing by a black hole, an infinite number of Einstein rings are formed by the light rays which wind around the black hole 
nearly on the photon sphere, which are 
called relativistic Einstein rings. This is also the case for the lensing by a wormhole.
In this paper, we study the Einstein ring and relativistic Einstein rings for the Schwarzschild black hole and the Ellis wormhole, the latter of which is an example of traversable wormholes of the Morris-Thorne class.
Given the configuration of the gravitational lensing and the radii
of the Einstein ring and relativistic Einstein rings, 
we can distinguish between 
a black hole and a wormhole in principle.
We conclude that 
we can detect the relativistic Einstein rings by wormholes which have the radii of the throat $a\simeq 0.5$pc at a galactic center with the distance $10$Mpc 
and which have $a\simeq 10$AU in our galaxy  
using by the most powerful modern instruments which have the resolution of $10^{-2}$arcsecond such as a $10$-meter optical-infrared telescope.
The black holes which make the Einstein rings of the same size as the ones by the wormholes are galactic supermassive black holes 
and the relativistic Einstein rings by the black holes are too small to measure with the current technology.
We may test the hypotheses of astrophysical wormholes by using the Einstein ring and relativistic Einstein rings in the future.  
       
\end{abstract}


\preprint{
RUP
-11-5
}
\thispagestyle{empty}

\maketitle

\section{I. ~Introduction}
Gravitational lensing is a very useful tool for astrophysics and cosmology. 
At first the gravitational lensing mainly was investigated on a theoretical basis in the weak gravitational field.
Using the gravitational lensing, 
we determine the cosmological constant, the distribution of dark matter and the Hubble constant, 
the existence of extrasolar planets and so on 
(see Schneider \textit{et al.} \cite{Gravitational_lenses} and Perlick \cite{Perlick_2004_Living_Rev,Perlick_2010} for the detail of the gravitational lens, and references therein). 

For the last decade, gravitational lensing in the strong gravitational field has been studied eagerly 
(see Virbhadra and Keeton \cite{Virbhadra_Keeton_2008}, Virbhadra \cite{Virbhadra_2009}, Bozza \cite{Bozza_2010}, Bozza and Mancini \cite{Bozza_Mancini_2012} and references therein). 
Frittelli \textit{et al.} \cite{Frittelli_Kling_Newman_2000}, 
Virbhadra and Ellis \cite{Virbhadra_Ellis_2000,Virbhadra_Ellis_2002} and 
Bozza \textit{et al.}  \cite{Bozza_Capozziello_Iovane_Scarpetta_2001}
studied the gravitational lensing in the strong field with the Schwarzschild spacetime 
and found the infinite Einstein rings 
which are too close to each other to separately resolve.
In this paper, we call these rings relativistic Einstein rings. 
The gravitational lensing in the strong field on the spherically symmetric static spacetime was investigated by 
Bozza \cite{Bozza_2002}, 
Hasse and Perlick \cite{Hasse_Perlick_2001} and 
Perlick \cite{Perlick_2004_Phys_Rev_D}. 
They showed that the relativistic Einstein rings 
are formed not only in the Schwarzschild spacetime but also in the other spherically symmetric static spacetime.

General relativity permits nontrivial topology of the spacetime  such as wormhole spacetimes 
(see Visser \cite{Lorentzuan_Wormholes} for the details of wormholes). 
Some hypotheses of astrophysical wormholes have been investigated \cite{Harko_Kovacs_Lobo_2009,Abdujabbarov_Ahmedov_2009,Pozanenko_Shatskiy_2010}.
For example, Kardashev \textit{et al.} suggest that some active galactic nuclei and other compact astrophysical objects may be explained as wormholes \cite{Kardashev_Novikov_Shatskiy_2007}.
We may test these hypotheses by using the gravitational lensing in the future.  

Kim and Cho \cite{Kim_Cho_1994} 
and Cramer \textit{et al.} \cite{Cramer_Forward_Morris_Visser_Benford_Landis_1995} 
pioneered gravitational lensing effects by wormholes.
Since then, the gravitational lensing effects by various wormholes have been investigated 
\cite{Rahaman_Kalam_Chakraborty_2007,Safonova_Torres_2002,Safonova_Torres_Romero_2002,Safonova_Torres_Romero_2001,Eiroa_Romero_Torres_2001,Nandi_Zhang_Zakharov_2006,Cramer_Forward_Morris_Visser_Benford_Landis_1995}. 

The Ellis spacetime which was investigated by Ellis \cite{Ellis_1973} 
is an example of traversable wormholes of the Morris-Thorne class \cite{Morris_Thorne_1988,Morris_Thorne_Yurtsever_1988}. 
The deflection angle of light in the Ellis wormhole geometry was studied by 
Chetouani and Clement \cite{Chetouani_Clement_1984} and recently Nakajima and Asada  \cite{Nakajima_Asada_2012}. 
The gravitational lensing on the Ellis geometry was studied by 
Dey and Sen \cite{Dey_Sen_2008}, Abe \cite{Abe_2010} and Toki \textit{et al.} \cite{Toki_Kitamura_Asada_Abe_2011} in the weak gravitational field 
and Perlick \cite{Perlick_2004_Phys_Rev_D}, Nandi \textit{et al.} \cite{Nandi_Zhang_Zakharov_2006} and Tejeiro and Larranaga \cite{Tejeiro_Larranaga_2005} in the strong gravitational field.

Perlick \cite{Perlick_2004_Phys_Rev_D}, Nandi \textit{et al.} \cite{Nandi_Zhang_Zakharov_2006} and Tejeiro and Larranaga \cite{Tejeiro_Larranaga_2005}
pointed out that the qualitative features of the gravitational lensing in the Ellis spacetime 
are very similar to the ones in the Schwarzschild spacetime for their photon spheres and their asymptotic flatness. 

In this paper, we will consider the Einstein ring and relativistic Einstein rings in the Ellis spacetime and the Schwarzschild spacetime, both of 
which are static and spherically symmetric ones.
We ask whether 
we can distinguish the Einstein-ring systems on the Schwarzschild spacetime and on the Ellis spacetime.
To answer this question, we focus on the relations between the Einstein ring and the relativistic Einstein rings.

This paper is organized as follows.
In Sec. II we will review the deflection angle on the Ellis wormhole spacetime. 
In Sec. III we give the radii of the Einstein ring and relativistic Einstein rings in the Ellis spacetime. 
In Sec. IV we will compare the Einstein ring and the relativistic Einstein rings in the Ellis spacetime to the ones in the Schwarzschild spacetime. 
In Sec. V we summarize and discuss our result.
In this paper we use the units in which $c=1$.

\section{II. ~Ellis wormhole spacetime and deflection angle}
In this section, we review the deflection angle on the Ellis wormhole spacetime \cite{Chetouani_Clement_1984,Nakajima_Asada_2012}. 
The line element in the Ellis wormhole solution is written in 
the following form:
\begin{equation}
ds^{2}=-dt^{2}+dr^{2}+(r^{2}+a^{2})d\Omega^{2},
\end{equation}
where $d\Omega^{2}=d\theta^{2}+\sin^{2}\theta d\phi^{2}$ and $a$ is a positive constant.
Introducing $\rho^{2}=r^{2}+a^{2}$, we can rewrite this into 
\begin{equation}
ds^{2}=-dt^{2}+\left(1-\frac{a^{2}}{\rho^{2}}\right)^{-1}d\rho^{2}+\rho^{2}d\Omega^{2},
\end{equation}
where $\rho=\pm a $ corresponds to the wormhole throat.
The spacetime has the Killing vectors $t^{\mu}\partial_{\mu}=\partial_{t}$ and  
$\phi^{\mu}\partial_{\mu}=\partial_{\phi}$ for stationarity and axial symmetry. 

We can concentrate ourselves on the equatorial plane 
because of spherical symmetry.
Using the conservation of the 
energy $E\equiv -g_{\mu \nu}k^{\mu}t^{\nu}$ 
and angular momentum $L\equiv g_{\mu \nu}k^{\mu}\phi^{\nu}$ 
and $k^{\mu}k_{\mu}=0$, where $k^{\mu}$ is the photon wave number, the photon trajectory is then given by 
\begin{equation}
\frac{1}{\rho^{4}}\left(\frac{d\rho}{d\phi}\right)^{2}
=\frac{1}{b^{2}}\left(1-\frac{a^{2}}{\rho^{2}}\right)
\left(1-\frac{b^{2}}{\rho^{2}}\right),
\end{equation}
where $b\equiv L/E$ is the impact parameter of the photon.
  
We can see that the photon is scattered if $|b|>a$, while it
reaches the throat if $|b|<a$. Since we are interested in the 
scattering problem, we assume $|b|>a$. 
Using $u=1/\rho$, 
we find 
\begin{equation}
\left(\frac{du}{d\phi}\right)^{2}=\frac{1}{b^{2}}(1-a^{2}u^{2})
(1-b^{2}u^{2}).
\end{equation}
Putting 
\begin{equation}
G(u)=a^{2}(a^{-2}-u^{2})(b^{-2}-u^{2}), 
\end{equation}
the azimuthal angle $\phi$
can be given as a function of $u$ by
\begin{equation}
\phi=\pm \int^{b^{-1}}_{u}\frac{du}{\sqrt{G(u)}}.
\end{equation}
Here we have set $\phi(b^{-1})=0$. 
The deflection angle $\alpha $ is then calculated to give
\begin{equation} 
\alpha=2\int^{b^{-1}}_{0}\frac{du}{\sqrt{G(u)}}-\pi.
\end{equation}

In the present case, we find 
\begin{equation}
\int^{b^{-1}}_{0}\frac{du}{\sqrt{G(u)}}=\int^{\pi/2}_{0}
\frac{d\theta}{\sqrt{1-\left(\frac{a}{b}\right)^{2}\sin^{2}\theta}}
=K\left(\frac{a}{b}\right), 
\end{equation}
where we have transformed $u=b^{-1}\sin \theta$
and $K(k)$ denotes the complete elliptic integral of the first 
kind (for example, see \cite{Handbook_of_Elliptic_Integrals_for_Engineers_and_Scientists}), 
which is defined as 
\begin{equation}\label{eq:complete_elliptic_integral}
K(k)=\int^{\pi/2}_{0}\frac{d\theta}{\sqrt{1-k^{2}\sin^{2}\theta}}.
\end{equation}
Hence, the deflection angle is given by
\begin{equation}
\alpha=2 K\left(\frac{a}{b}\right)-\pi.
\end{equation}
Since $K(k)$ admits a power series
\begin{equation}
K(k)=\frac{\pi}{2}\sum_{n=0}^{\infty}\left[\frac{(2n-1)!!}{(2n)!!}\right]^{2}k^{2n},
\end{equation}
where $n!!$ denotes the double factorial of $n$ and $(-1)!!=1$,  
we get the deflection angle 
\begin{equation}
\alpha=
\pi \sum_{n=1}^{\infty}
\left[\frac{(2n-1)!!}{(2n)!!}\right]^{2}\left(\frac{a}{b}\right)^{2n}.
\end{equation}
Thus, the deflection angle is approximately given in the 
weak-field regime $|b|\gg a$ by
\begin{equation} \label{eq:weak_field_deflection_angle}
\alpha \simeq \frac{\pi}{4}\left(\frac{a}{b}\right)^{2}.
\end{equation}
In general, the deflection angle is always greater than its weak-field approximation
and is diverging as $|b|\to a$.

\section{III. ~Einstein ring and relativistic Einstein rings of Ellis wormhole}
In this section, we examine the diameter angles of the Einstein ring and the relativistic Einstein rings on the Ellis spacetime.
Now we will consider the case that both the observer and the source object are far from the lensing object, 
or $D_{l}\gg b$ and $D_{ls}\gg b$, 
where $D_{l}$ and $D_{ls}$ are the separations between the observer and 
lens and between the lens and source, respectively.
The configuration of the gravitational lensing is given in Fig. 1.
\begin{figure}[htbp]
 \begin{center}
  \includegraphics[width=80mm]{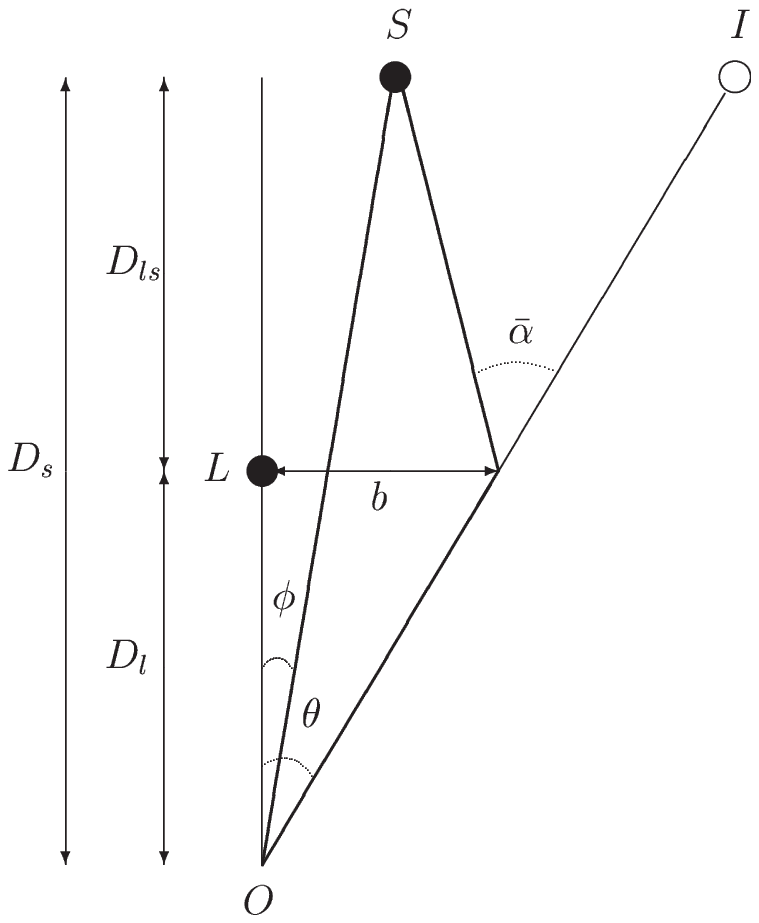}
 \end{center}
 \caption{The configuration of the gravitational lensing.
The light rays emitted by the source $S$ are deflected 
by the lens $L$ (a wormhole or a black hole) 
and reach the observer $O$ with the angle of the lensed image $\theta$, 
instead of the real angle $\phi$. 
$b$ and $\bar{\alpha}$ are the impact parameter and the effective deflection angle, respectively. 
$D_{l}$ and $D_{ls}$ are the separations between the observer and 
the lens and between the lens and the source, respectively.}
 \label{fig:one}
\end{figure}
Then, the lens equation is given by 
\begin{equation}\label{eq:lens_equation}
D_{ls}\bar{\alpha}=D_{s}(\theta-\phi),
\end{equation}
where $\bar{\alpha}=(\alpha~\mbox{mod}~2\pi)$ is the effective deflection angle, 
$\theta$ and $\phi$ are the angles of the lensed image and
the real image from the observer, respectively, 
and $D_{s}=D_{l}+D_{ls}$ is the separation between the observer and source.
Note that we have assumed $|\bar{\alpha}|\ll 1$, $|\theta|\ll 1$ and $|\phi|\ll 1$.
The deflection angle can be expressed $\alpha = \bar{\alpha} + 2\pi n$,
where $n$ is a non-negative integer, denoting the winding number of the light ray. 

The ring image corresponds to the image angle $\theta$ for vanishing real angle $\phi=0$.
By the symmetry, the image is necessarily a ring with the diameter angle $\theta$.

Since $b=D_{l}\theta$, we find that the ring image is given by 
\begin{equation}
\theta_{n}=\frac{a}{D_{l}}\frac{1}{k_{n}},
\end{equation}
where $k_{n}\in (0,1)$ is a unique root of the transcendental equation
\begin{eqnarray}
2K(k)-\frac{\eta}{k}&=&\left(2n+1\right)\pi, 
\label{eq:transcendental}\\
\eta&=&\frac{D_{s}}{D_{l}D_{ls}}a.
\end{eqnarray}
We should note that $2K(k)-\eta/k$ is monotonically increasing with respect to $k$ and changes from $-\infty $ to $\infty$ as $k$ increases from $0$ to $1$.
The uniqueness of the root follows from the monotonicity. 
Moreover, we can conclude that $k_{n}$ monotonically increases and approaches 
1 as $n\to \infty$
and hence the image angle $\theta_{n}$ monotonically decreases and 
approaches $a/D_{l}$.

In the weak-field regime $|b|\gg a$, the winding number $n$ should be $n=0$.
Using the deflection angle (\ref{eq:weak_field_deflection_angle}), 
we can solve the equation (\ref{eq:transcendental}) approximately and get 
the diameter angle of the Einstein ring
\begin{eqnarray}\label{eq:worm_einstein_ring}
\theta_{0}&&\simeq  \left( \frac{\pi}{4} \frac{D_{ls}}{D_{s}D_{l}^{2}}a^{2} \right)^\frac{1}{3} \nonumber\\
&&\simeq 2.0 \, \textrm{arcsecond} \left(\frac{D_{ls}}{10 \textrm{Mpc}} \right)^{\frac{1}{3}} \left(\frac{20 \textrm{Mpc}}{D_{s}} \right)^{\frac{1}{3}} \left(\frac{10 \textrm{Mpc}}{D_{l}} \right)^{\frac{2}{3}} \left(\frac{a}{0.5\textrm{pc}} \right)^{\frac{2}{3}}. \qquad 
\end{eqnarray}
This approximation is good for $D_{l}\gg a$ and $D_{ls}\gg a$.
The relative error is $\sim 10^{-2}$ for $a=0.5$pc and $D_{l}=D_{ls}=10$ Mpc. 

In the especially strong-field regime, 
where the winding number $n$ becomes $n\geq 1$, we can easily check 
that $a \simeq b$ or $k_{n}\simeq 1$ satisfies the transcendental equation (\ref{eq:transcendental}) in numerical calculations.
Physically this means that the light rays which wind around the wormhole 
nearly on the photon sphere make the relativistic Einstein rings \cite{Bozza_2002,Perlick_2004_Phys_Rev_D}.  
Then the diameter angles of the relativistic Einstein rings are approximately given by 
\begin{eqnarray}\label{eq:worm_relativistic_einstein_ring}
\theta_{n\ge 1} &\simeq & \frac{a}{D_{l}}\nonumber\\
&\simeq&1.0\times 10^{-2} \, \textrm{arcsecond} 
\left(\frac{10 \textrm{Mpc}}{D_{l}} \right) \left(\frac{a}{0.5\textrm{pc}} \right). \qquad
\label{eq:theta_n_wormhole} 
\end{eqnarray}
Regardless of the values of $D_{ls}$, $D_{l}$ and $a$, 
the relative error of the above approximation to the direct numerical solution of the outermost relativistic Einstein ring ($n=1$) is $\sim 10^{-3}$ 
and those of the other relativistic Einstein rings ($n\ge 2$) are smaller than $10^{-5}$.
This implies that it is difficult to resolve each relativistic Einstein ring separately.

Thus, we conclude that there is one Einstein-ring image and countably infinite relativistic Einstein-ring images, the latter of which  
accumulate to form the apparently single ring image of the throat with 
the diameter $a/D_{l}$. This conclusion does not depend on the 
value of $\eta$.

If we are given the distance $D_{s}$ to the source from the observer, 
the distance $D_{l}$ to the lens from the observer 
and the radius $\theta_{0}$ of the Einstein ring, 
we can determine the radius of the throat $a$  
from Eq. (\ref{eq:worm_einstein_ring}). 
Then, we can use $\theta_{n\ge 1}$ (\ref{eq:worm_relativistic_einstein_ring}) to test the assumption that the lens object is a wormhole.

From Eqs. (\ref{eq:worm_einstein_ring}) and (\ref{eq:worm_relativistic_einstein_ring}) we obtain the relation between $\theta_{0}$ and $\theta_{n\ge 1}$ by
\begin{eqnarray}\label{eq:thata_ralation_Ellis}
\theta_{n\ge 1}\simeq \left( \frac{4}{\pi}\frac{D_{s}}{D_{ls}}\right)^{\frac{1}{2}}\theta_{0}^{\frac{3}{2}}. 
\label{eq:theta_0_n_wormhole}
\end{eqnarray}
This relation generally holds in astrophysical situations, as long as 
$a \ll D_{l}$ and $a\ll D_{ls}$ are satisfied.
If the lens is identified with a wormhole, 
we can even estimate the radius of the wormhole throat in terms of $D_{s}$, $\theta_{0}$
and $\theta_{n}$ through
\begin{equation}
a \simeq \theta_{n\ge 1}\left(1-\frac{4}{\pi}\theta_{n\ge 1}^{-2}\theta_{0}^{3}\right)D_{s}.
\end{equation}  
This follows from Eqs.~(\ref{eq:theta_n_wormhole}) and (\ref{eq:theta_0_n_wormhole}).

\section{IV. ~Comparison between wormholes and black holes}
In this section we compare the Einstein ring and the relativistic Einstein rings in the Ellis spacetime to the ones in the Schwarzschild spacetime 
and show that we can distinguish between black holes and wormholes. 

Now, we will briefly review the deflection angle, Einstein ring and relativistic Einstein rings 
for the Schwarzschild spacetime \cite{Bisnovatyi_Tsupko_2008,Bozza_Capozziello_Iovane_Scarpetta_2001,Muller_2008,Virbhadra_Ellis_2000,Virbhadra_Ellis_2002,Hasse_Perlick_2001}
and present the relation between $\theta_{0}$ and $\theta_{n\ge 1}$.

In the weak-field regime $b\gg r_{g}$, where $r_{g}=2GM$ is the Schwarzschild radius of the black hole of mass $M$, 
the deflection angle is approximately given by
\begin{equation} \label{eq:weak_field_deflection_angle_black_hole}
\alpha \simeq \frac{2r_{g}}{b}.
\end{equation}
The diameter angle of the Einstein ring is given by  
\begin{eqnarray}\label{eq:black_einstein_ring}
\theta_{0} &&\simeq \sqrt{\frac{2D_{ls}}{D_{l}D_{s}}r_{g}} \nonumber\\
&&\simeq 2.0 \, \textrm{arcsecond} \left(\frac{D_{ls}}{10 \textrm{Mpc}} \right)^{\frac{1}{2}} \left(\frac{M}{10^{10}M_{\odot}} \right)^{\frac{1}{2}} \left(\frac{10 \textrm{Mpc}}{D_{l}} \right)^{\frac{1}{2}} \left(\frac{20 \textrm{Mpc}}{D_{s}} \right)^{\frac{1}{2}}. \qquad 
\label{eq:theta_0_blackhole}
\end{eqnarray}
We can determine $r_{g}$ in the same way as the radius of the throat $a$.

In the especially strong-field regime, where the winding number $n$ becomes $n\geq 1$,
the impact parameter $b$ that satisfies the lens equation  
should be nearly the critical impact parameter 
$b \simeq (3\sqrt{3}/2)r_{g}$ (see \cite{Bozza_Capozziello_Iovane_Scarpetta_2001,Bozza_2002,Virbhadra_Ellis_2000}).
Then the diameter angles of the inseparable 
relativistic Einstein rings $\theta_{n\ge 1}$ are given by 
\begin{eqnarray}
\theta_{n\ge 1} &\simeq & \frac{3\sqrt{3}}{2}\frac{r_{g}}{D_{l}} \nonumber\\
&\simeq &5.1\times 10^{-5} \, \textrm{arcsecond} \left(\frac{M}{10^{10}M_{\odot}} \right) \left(\frac{10 \textrm{Mpc}}{D_{l}} \right). \qquad 
\label{eq:theta_n_blackhole}
\end{eqnarray}

It is useful to remember that the leading term of the deflection angle in the weak-field regime 
is the second order of the small amount $a/b$ on the Ellis geometry (\ref{eq:weak_field_deflection_angle}), 
while it is the first order of the small amount $r_{g}/b$ on the Schwarzschild geometry (\ref{eq:weak_field_deflection_angle_black_hole}). 
So the relation between $\theta_{0}$ and $\theta_{n\ge 1}$ for the Schwarzschild spacetime 
\begin{eqnarray}\label{eq:relation_diameter_angle_black_hole}
\theta_{n\ge 1}\simeq \frac{3\sqrt{3}}{4}\frac{D_{s}}{D_{ls}}\theta_{0}^{2} 
\label{eq:theta_n_0_blackhole}
\end{eqnarray}
is different from that on the Ellis spacetime (\ref{eq:thata_ralation_Ellis}).
Figure. 2 shows the angle of the relativistic Einstein ring $\theta_{n\ge 1}$ versus the angle of the Einstein ring $\theta_{0}$ for $D_{l}=D_{ls}=10$Mpc.
Thus, we can distinguish between black holes and wormholes in principle 
if we are given $D_{s}$, $D_{l}$, $\theta_{0}$ and $\theta_{n\ge 1}$.
\begin{figure}[htbp]
 \begin{center}
  \includegraphics[width=87mm]{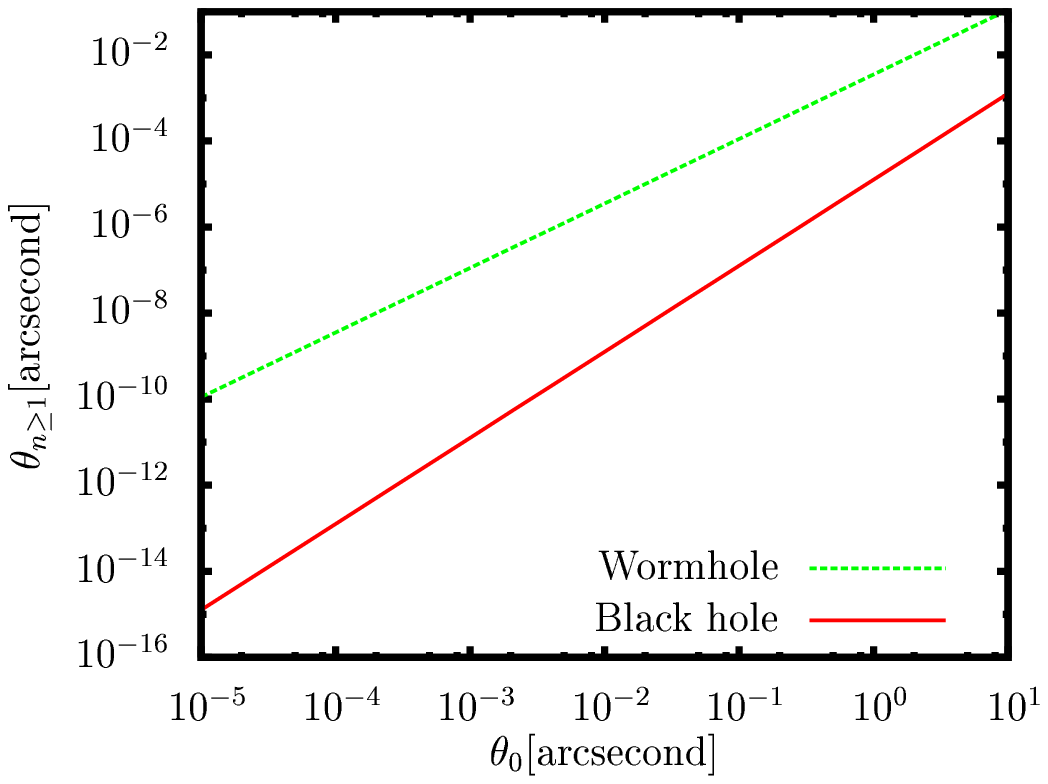}
 \end{center}
 \caption{The angle of the relativistic Einstein ring $\theta_{n\ge 1}$ versus the angle of the Einstein ring $\theta_{0}$ for $D_{l}=D_{ls}=10$Mpc.
 The broken (green) and solid (red) lines plot the cases where the lens objects are a wormhole and a black hole, respectively.
 }
 \label{fig:two}
\end{figure}

If the lens is identified with a black hole, we can estimate the 
Schwarzschild radius and,hence, the black hole mass by 
\begin{equation}
r_{g}\simeq \frac{2}{3\sqrt{3}}\theta_{n\ge 1}\left(1-\frac{3\sqrt{3}}{4}\theta_{0}^{2}\theta_{n\ge 1}^{-1}\right)D_{s}
\end{equation}
in terms of $\theta_{0}$, $\theta_{n\ge 1}$ and $D_{s}$. This follows from Eqs.~(\ref{eq:theta_n_blackhole}) and (\ref{eq:theta_n_0_blackhole}).

\section{V. ~Discussion and conclusion}
It is well known that
the qualitative features of the gravitational lensing on the Ellis spacetime 
are very similar to the ones on the Schwarzschild spacetime for their photon spheres and their asymptotic flatness \cite{Perlick_2004_Phys_Rev_D,Nandi_Zhang_Zakharov_2006,Tejeiro_Larranaga_2005}.
However, we realize that their quantitative features are very different due to their different weak-field behaviors.

We consider the experimental situation where we know the 
separation $D_{s}$ between the observer and the source 
and the separation $D_{l}$ between the observer and the lens.
We assume that we do not know whether the lens object is a black bole or a wormhole
and do not know its parameter, i.e., the mass $M$ or the radius $a$ of the throat in advance.

We need at least two observable quantities to determine whether the lens object is a black hole or wormhole 
since the lens system has one parameter in this situation.
First, we observe an Einstein ring and determine the parameter for both possibilities.
Second, we observe relativistic Einstein rings 
and tell the wormhole from the black hole.
If the predicted relativistic ring angles by the black hole and by the wormhole were of similar size, we could not discern the difference.
However, Eqs. (\ref{eq:thata_ralation_Ellis}), (\ref{eq:relation_diameter_angle_black_hole}) and Fig. \ref{fig:two}
show that we do not confuse them.

We conclude that we can detect the relativistic Einstein rings by wormholes which have $a\simeq 0.5$pc at a galactic center with the distance $D_{l}=D_{ls}=10$Mpc 
and which have $a\simeq 10$AU in our galaxy with the distance $D_{l}=D_{ls}=10$kpc 
using the most powerful modern instruments which have the resolution of $10^{-2}$arcsecond such as a $10$-meter optical-infrared telescope.
Note that the corresponding black holes which have the Einstein rings of the same size are galactic supermassive black holes with $10^{10}M_{\odot}$ and $10^{7}M_{\odot}$, respectively, 
and that the relativistic Einstein rings by these black holes are too small to measure with the current technology.

In fact, our results imply that we can distinguish between slowly rotating Kerr-Newmann black holes
and the Ellis wormholes with their Einstein-ring systems. This is because the leading
term of the deflection angle for the lensing by the Kerr-Newmann black holes in the weak-field regime
is equal to the one for the lensing by the Schwarzschild black holes, while the black hole charge and
small spin only slightly change the radii of the relativistic Einstein rings \cite{Bozza_2002,Eiroa_Romero_Torres_2002,Sereno_2004,Bozza_2003,Sereno_Luca_2006}.
Moreover, this also suggests that it is much more challenging to determine the charge and/or small
spin of black holes than to distinguish between black holes and the Ellis wormholes.

We assumed that the observer, the lensing object and the source object are directly-aligned,  
though such a  configuration is fairly rare.
In general the strong gravitational lensing effect is observed as broken-ring images which are called relativistic images \cite{Virbhadra_Ellis_2000}.
Therefore, more realistic problem is to size the relativistic images.
Our result suggests that we can distinguish black holes and wormholes by using the relativistic images.
To observe the relativistic images is one of the challenging works with many difficulties. 
Bozza \textit{et al.} pointed out that relativistic images are always very faint with respect to the weak field images \cite{Bozza_Capozziello_Iovane_Scarpetta_2001}.
The Very Large Telescope Interferometer (VLTI) has high resolution \cite{Delplancke_Gorski_Richichi_2001,Eckart_Bertram_et_al_2003}
but it will not work because of this demagnifying effect.

We also assumed point-like sources,  
although astrophysical sources have their own size.
If the source object is a galaxy, it may conceal the relativistic Einstein rings, 
especially, in the case that the lens object is a black hole.
Testing some hypotheses of astrophysical wormholes by using the relativistic Einstein rings and the Einstein ring is left as future work.  

Tejeiro and Larranaga \cite{Tejeiro_Larranaga_2005} investigated 
the gravitational lensing effect of the wormhole solution obtained by connecting the Ellis solution as an interior region and the Schwarzschild solution as an exterior region \cite{Lemos_Lobo_Quinet_2003}.
They concluded that we cannot distinguish the Schwarzschild black hole and the wormhole 
unless the discontinuity of the magnification curve at the boundary is observed.
This does not contradict our results 
because their wormhole solution behaves as the Schwarzschild solution in the weak-field regime.
\\
\\
{\bf Acknowledgements:}
\\

The authors would like to thank 
U. Miyamoto, H. Asada and  F. Abe 
for valuable comments and discussion. 
The work of N. T. and K. Y. was supported in part
by Rikkyo University Special Fund for Research. 
T. H. would also thank the Astronomy
Unit, Queen Mary, University of London for its hospitality.
T. H. was supported by the Grant-in-Aid for Young Scientists (B) (No. 21740190)
and the Grant-in-Aid for Challenging Exploratory Research (No. 23654082)
for Scientific
Research Fund of the Ministry of Education, Culture, Sports, Science
and Technology, Japan.
N. T. thanks the Yukawa Institute for Theoretical Physics at Kyoto University, 
where this work progressed during the YITP workshop YITP-W-12-08 on "Summer School on Astronomy \& Astrophysics 2012".

\end{document}